\documentclass[]{aa}
\usepackage{natbib}
\usepackage{lineno}
\usepackage{amsmath}
\usepackage{fixltx2e}
\usepackage[usenames,dvipsnames]{color}
\usepackage{multirow}
\usepackage{graphicx}
\usepackage{hyperref}
\bibpunct{(}{)}{;}{a}{}{,}

\newcommand{\fermi}{${\it Fermi}$}
\renewcommand{\figurename}{Figure}

\begin{document}

\title{The first full orbit of Eta Carinae seen by Fermi}
\author{
K.~Reitberger$^{(1)}$ \and
A.~Reimer$^{(1,2)}$ \and
O.~Reimer$^{(1,2)}$ \and
H.~Takahashi$^{(3)}$
}
\authorrunning{LAT collaboration}

\institute{
\inst{1}~Institut f\"ur Astro- und Teilchenphysik und Institut f\"ur Theoretische Physik, Leopold-Franzens-Universit\"at Innsbruck, A-6020 Innsbruck, Austria\\
\inst{2}~Kavli Institute for Particle Astrophysics and Cosmology, Department of Physics and SLAC National Accelerator Laboratory, Stanford University, Stanford, CA 94305, USA\\
\inst{3}~Department of Physical Sciences, Hiroshima University, Higashi-Hiroshima, Hiroshima 739-8526, Japan\\
\email{klaus.reitberger@uibk.ac.at} \\
}
\date{Received ... ; accepted ...}

\abstract
{}
{The binary system $\eta$ Carinae has completed its first 5.54~y orbit since the beginning of science operation of the \textit{Fermi} Large Area Telescope (LAT). We are now able to investigate the high-energy $\gamma$-ray source at the position of $\eta$ Carinae over its full orbital period. By this, we can address and confirm earlier predictions for temporal and spectral variability.}
{Newer versions of the LAT datasets, instrument response functions and background models allow for a more accurate analysis. Therefore it is important to re-evaluate the previously analyzed time period along with the new data to further constrain location, spectral shape, and flux time history of the $\gamma$-ray source.}
{We confirm earlier predictions of increasing flux values above 10 GeV toward the next periastron passage. For the most recent part of the data sample, flux values as high as those before
the first periastron passage in 2008 are recorded. A comparison of spectral energy distributions around periastron and apastron passages reveals strong variation in the high-energy band. This
is due to a second spectral component that is present only around periastron.}
{Improved spatial consistency with the $\gamma$-ray source at the position of $\eta$ Carinae along with the confirmation of temporal variability above 10 GeV in conjunction with the orbital period strengthens the argument for unambiguous source identification. Spectral variability provides additional constraints for future modeling of the particle acceleration and $\gamma$-ray emission in colliding-wind binary systems.}{}


 \maketitle

\section{Introduction}

Five years have passed since the first reports of the unexpected presence of a strong high-energy $\gamma$-ray source at the position of the enigmatic binary system $\eta$ Carinae \citep{Agile,Takapaper}.
This binary system is located 2.3 kpc away from Earth and comprises a massive LBV star and an O- or B-type companion star in an highly eccentric orbit ($e \sim0.9$). Since the first reports, several studies have tried to further investigate the system's nonthermal emission. Subsequent analyses with the \textit{Fermi}-LAT \citep{Walter, Reitberger2012} indicated flux variability in concordance with the orbital period of the binary system. Although the source was detected up to energies of 300 GeV with \textit{Fermi}-LAT, ground-based Imaging Atmospheric Cherenkov Telescopes (IACT) did not yet report a detection \citep{Hess}. Below the MeV range, evidence remains for a nonthermal hard X-ray component \citep{Leyder,Sekiguchi}, which is presumably linked to the emission as seen with \textit{Fermi}-LAT.

Apart from alternative suggestions \citep[e.g., ][]{Ohm}, the most common interpretation of $\gamma$-ray emission in the $\eta$ Carinae binary system is a colliding-wind binary scenario. In this, charged particles are accelerated at the shock fronts of an extensive wind-collision region (WCR), which forms where the powerful stellar winds of the high-mass stars collide \citep{Eichler1993,Dougherty2006}. Traveling alongside the hot, shocked gas in the WCR, the charged particles (leptonic and hadronic in nature) interact with stellar radiation fields, magnetic fields, and the surrounding plasma, subsequently emitting nonthermal radiation via inverse Compton emission, synchrotron emission, relativistic bremsstrahlung, and neutral pion decay \citep{Reimer2006,Bednarek2011}.
Because the properties of the WCR change drastically with stellar separation, the ensuing nonthermal emission is expected to show strong modulation on orbital timescales for systems with high eccentricity such as $\eta$ Carinae.

With the conclusion of $\eta$ Carinae's first full orbit in \textit{Fermi}-LAT $\gamma$-ray data and improved analysis performance, it is now time to perform a dedicated data analysis to determine whether previous indications and predictions for flux variability can be confirmed over one orbital period in $\gamma$-rays.

\section{Observations with the Large Area Telescope }
\subsection{Dataset}
\label{data}
The time range of the \textit{Fermi}-LAT dataset (P7 reprocessed SOURCE class) that is used in this analysis spans 2024 days from 2008 August 4 to 2014 February 18 (Mission Elapsed Time: 239557417~s to 414431017~s). During this period, the position of $\eta$ Carinae has been monitored for a total of 399 days (at energies 0.2 to 10 GeV) and 392 days (at energies 10 to 300 GeV). Different exposures for different energy bands are due to the energy-dependent instrument response functions (IRFs). To account for irregular exposure due to targeted observations and other deviations of the prevalent regular sky-coverage of the \fermi-LAT, we considered bins of equal exposure time rather than real time in our temporal analysis.
The data were reduced and analyzed using the \textsc{Fermi Science Tools v9r31} package\footnote{See the Fermi Science Support Centre (FSSC) website for details:
\url{http://fermi.gsfc.nasa.gov/ssc/}}. To reject atmospheric $\gamma$\ rays from the Earth's limb, time periods when the region was observed at zenith angle greater than 105$^{\circ}$ were excluded and a rocking angle cut of less then 52$^{\circ}$ was applied.
We used all remaining photons with energy \mbox{E$\,>\,$200\,MeV} within a square of 21$^{\circ}$ base length (aligned in directions of right ascension $\alpha$ and declination $\delta$) centered on the nominal location of $\eta$ Car, ($\alpha$, $\delta$)\,=\,(161.265$^\circ$, $-$59.685$^\circ$).

\subsection{Likelihood analysis}
Because  $\eta$~Carinae is located at low Galactic latitude within the projection of the Carina-Sagittarius Arm of the Milky Way, the diffuse Galactic $\gamma$-ray background significantly affects the detected $\gamma$-ray flux and needs to be modeled with care. The size of the \fermi-LAT point-spread function (PSF) makes
knowing the neighboring $\gamma$-ray sources and their contributions equally important. The 68\% containment radius corresponds to more than 2$^{\circ}$ at 200 MeV (on-axis).

We performed maximum-likelihood analyses using the P7REP$\_$SOURCE$\_$V15 IRFs in conjunction with the Galactic diffuse model
\emph{template$\_$4years$\_$P7$\_$v15$\_$repro$\_$v3.fits} and the isotropic background component \emph{iso$\_$source$\_$v05.txt}. The applied source
model for all subsequent investigations includes 199 point sources that are located within a radius of 20$^{\circ}$ around the source of interest and are part of the \fermi-LAT 4-year (3FGL) catalog \citep{3FGL}. For each source we used the same spectral model as in the catalogue. The spectral parameters of all the sources within a radius of 10$^{\circ}$ were allowed to vary in the likelihood analysis. Most sources were modeled either by a simple power-law (PL)
\begin{equation}
\frac{dN}{dE}=A\:\left( \frac{E}{E_0}\right)^{-\alpha},
\end{equation} where $A$ is the normalization parameter and $\alpha$ is the spectral index, or by a LogParabola function, which provides additional information on the spectral curvature of some sources by introducing a break energy $E_b$ and the parameters $\alpha$ and $\beta$
 \begin{equation} \frac{dN}{dE}=A\left(\frac{E}{E_b}\right)^{-(\alpha+\beta \log(\frac{E}{E_b}))}. \end{equation}
Exceptions are $\gamma$-ray pulsars, which were modeled by a cutoff power-law (CPL) function expression
\begin{equation} \frac{dN}{dE}=B\: \left( \frac{E}{E_0}\right)^{-\beta}\mathrm{exp}\left(-\frac{E}{E_{\mathrm{b}}}\right), \end{equation}
where $B$ is the normalization, $\beta$ is the spectral index and $E_{\mathrm{b}}$ is the cutoff energy. In all three cases, $\frac{dN}{dE}$ gives the differential flux.\\

Source detection significance can be described by the likelihood test statistic value $TS=-2\ln(L_{\max,0}/L_{\max,1}),$ which compares the ratio of two values that are obtained by a maximum-likelihood procedure. $L_{\max,0}$ is the likelihood for a model without an additional source at a specified location (the null-hypothesis), and $L_{\max,1}$ is the maximum-likelihood value for a model including an additional source. Surrounding sources and background are taken into account in both cases. The same notion can also be applied to comparing the likelihood of two source models that are described by different spectral models for a specific source of interest. The corresponding TS-value can be converted into detection significance $\sigma$ for a point source via $\sigma\simeq\sqrt{TS}$ \citep{Mattox}.

\section{Analysis results}

\subsection{Spatial analysis}
\begin{figure*}[t]

\begin{minipage}{0.5\textwidth}
\centering
\includegraphics[width=\textwidth]{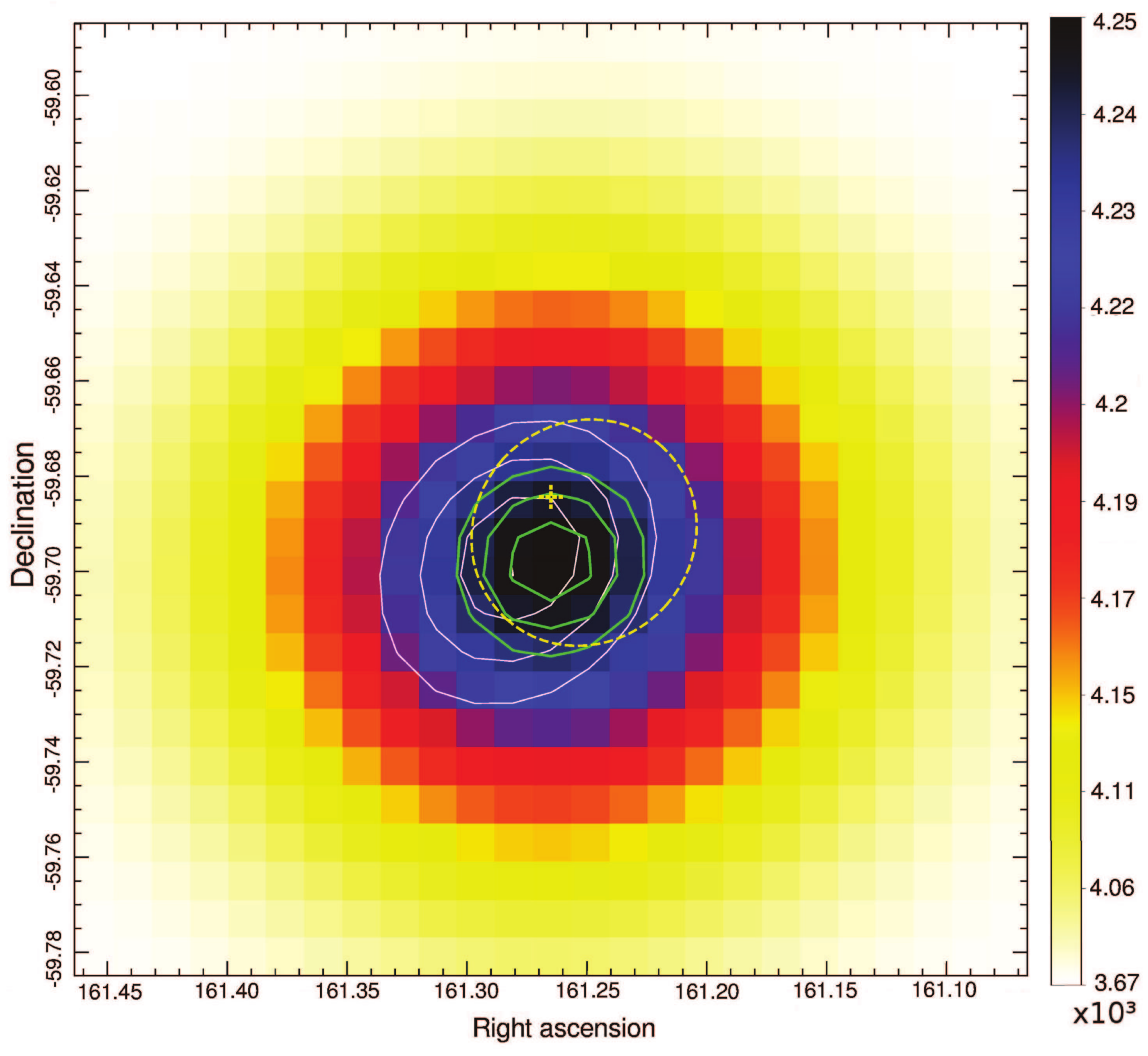}
\end{minipage}
\begin{minipage}{0.5\textwidth}
\centering
\includegraphics[width=\textwidth]{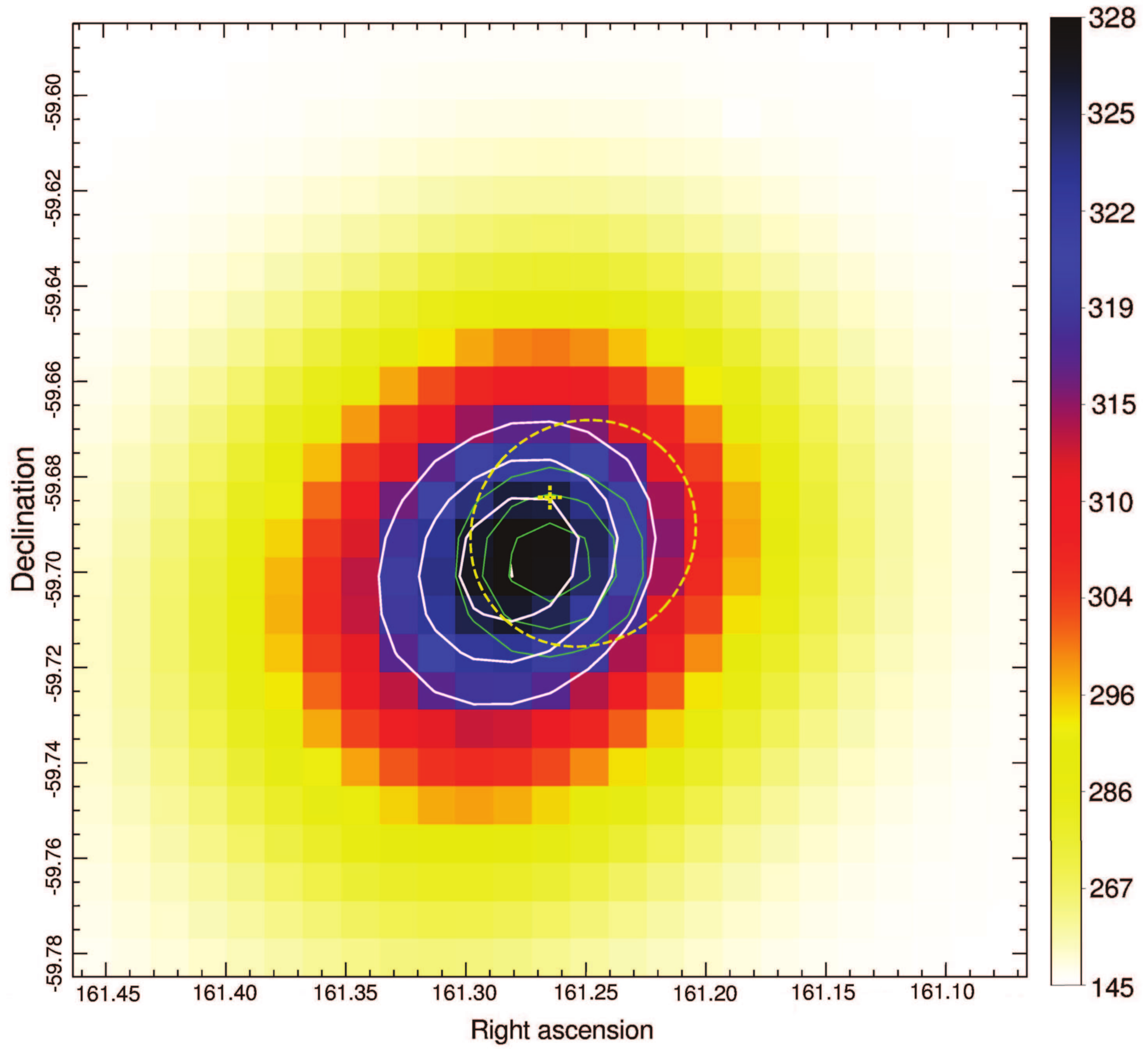}
\end{minipage}

\caption{Test statistics maps of the celestial region around 3FGL~J1045.15941 as obtained by the likelihood analysis. The TS-value corresponds to the
likelihood of the source being located in a specific grid point. Left plot: 0.2 to 10 GeV, right plot: 10 to 300 GeV. The confidence contours (green for the low band, white for the high band) mark the 68.3\%, 95.4\%, and 99.7\% uncertainty regions for the location of the source of interest.
The dashed yellow circle marks the cataloged 95\% source location region of 3FGL~J1045.15941. The nominal position of $\eta$~Carinae is indicated by the yellow cross. \label{L} }
\end{figure*}
In the two celestial coordinate grids in Fig. \ref{L}, the TS-values indicate the detection significance for a trial additional point source (a PL with two free parameters) that was considered sequentially at each of the grid positions.
The left figure covers the region around 3FGL~J1045.15941 (the name of the $\gamma$-ray source that is now associated with $\eta$ Carinae according to the 3FGL catalog \citep{3FGL}) at energies from 0.2 to 10 GeV, the right figure at energies from 10 to 300 GeV, each representing 2024 days of data. The energy band was divided according to the previous indication of two different emission components. The two TS-maps were obtained with the tool \textit{gttsmap} with $\eta$~Carinae removed from the source model. In addition, confidence contours mark the 68.3\%, 95.4\%, and 99.7\% uncertainty regions for the location of the source defined by the maximum TS value in the grids. In both energy bands the nominal position of $\eta$~Carinae is found within the 95.4\% error region. For the low band it is at the border of the the 68.3\% error region. 

In the low-energy band the $\gamma$-ray signal is centered on ($\alpha$, $\delta$)\,=\,(161.265$^\circ$, $-$59.698$^\circ$) with a fairly circular 95.4\% confidence region of radius $r$\,=\,0.014$^\circ$. In the high-energy band the signal is centered on a ($\alpha$, $\delta$)\,=\,(161.278$^\circ$, $-$59.698$^\circ$) with  with a fairly circular 95.4\% confidence region of radius $r$\,=\,0.021$^\circ$. Statistical limitations toward higher energies yield a larger confidence region for the hard band despite the larger PSF at lower energies.
We find that localization uncertainties (in terms of error contour radii) have decreased with respect to an earlier analysis \citep{Reitberger2012} by a factor of 1.6 for the low band and 2.3 for the high band. At the same time, the angular separation of the best-fit positions for the energy bands has decreased from 0.033$^\circ$ to merely 0.007$^\circ$.

The difference to the indicated catalog position of 3FGL~J1045.15941 is understood by acknowledging the difference in exposure and the decomposition into low- and high-energy bands.

Comparing these results to an earlier localization study with 35 months of data \citep{Reitberger2012}, spatial agreement in both energy bands is considerably improved (reduced error contours). The reason is not only the better statistics, but also the continuously improving IRFs and knowledge about residual backgrounds. The spatial agreement between the detected $\gamma$-ray emission and location of $\eta$ Car, as well as the agreement of the $\gamma$-ray source positions for the consecutive energy bands, is better than it has been for fewer datasets.

\label{flux}
Performing a maximum-likelihood fit using the tool \textit{gtlike} (dataset and source model as described above) with the source of interest fixed at the nominal position of $\eta$ Car, we determined the total $\gamma$-ray flux values for the two respective energy bands. The source of interest was modeled with a CPL for the low-energy band (0.2 to 10 GeV) and a PL in for the high-energy band 10 to 300 GeV). Systematic uncertainties were estimated using bracketing IRFs that take into account systematic variations of the \textit{Fermi}-LAT effective area \citep[see ][]{IRFpaper}.
At the low-energy band, the source has a flux of F$^{>\mathrm{200MeV}}_{<\mathrm{10GeV}}$\,=\,(1.00 $\pm$ 0.05$_{\mathrm{stat.}}$ $^{+0.12}_{-0.13}$ syst.)
$\times$ 10$^{-7}$
cm$^{-2}$s$^{-1}$ and a TS value of 3660 ($\simeq$ 60$\sigma$).

At the high-energy band, the source has a flux of F$^{>\mathrm{10GeV}}_{<\mathrm{300GeV}}$\,=\,(6.03 $\pm$ $^{+0.71}_{-0.67}$ stat. $^{+0.43}_{-0.37}$ syst.)
$\times$ 10$^{-10}$
cm$^{-2}$s$^{-1}$ and a TS value of 286 ($\simeq$ 17$\sigma$).

\subsection{Spectral analysis}
\label{SA}
\begin{figure*}[t]
\centering
\begin{minipage}{0.49\textwidth}
\includegraphics[width=\textwidth]{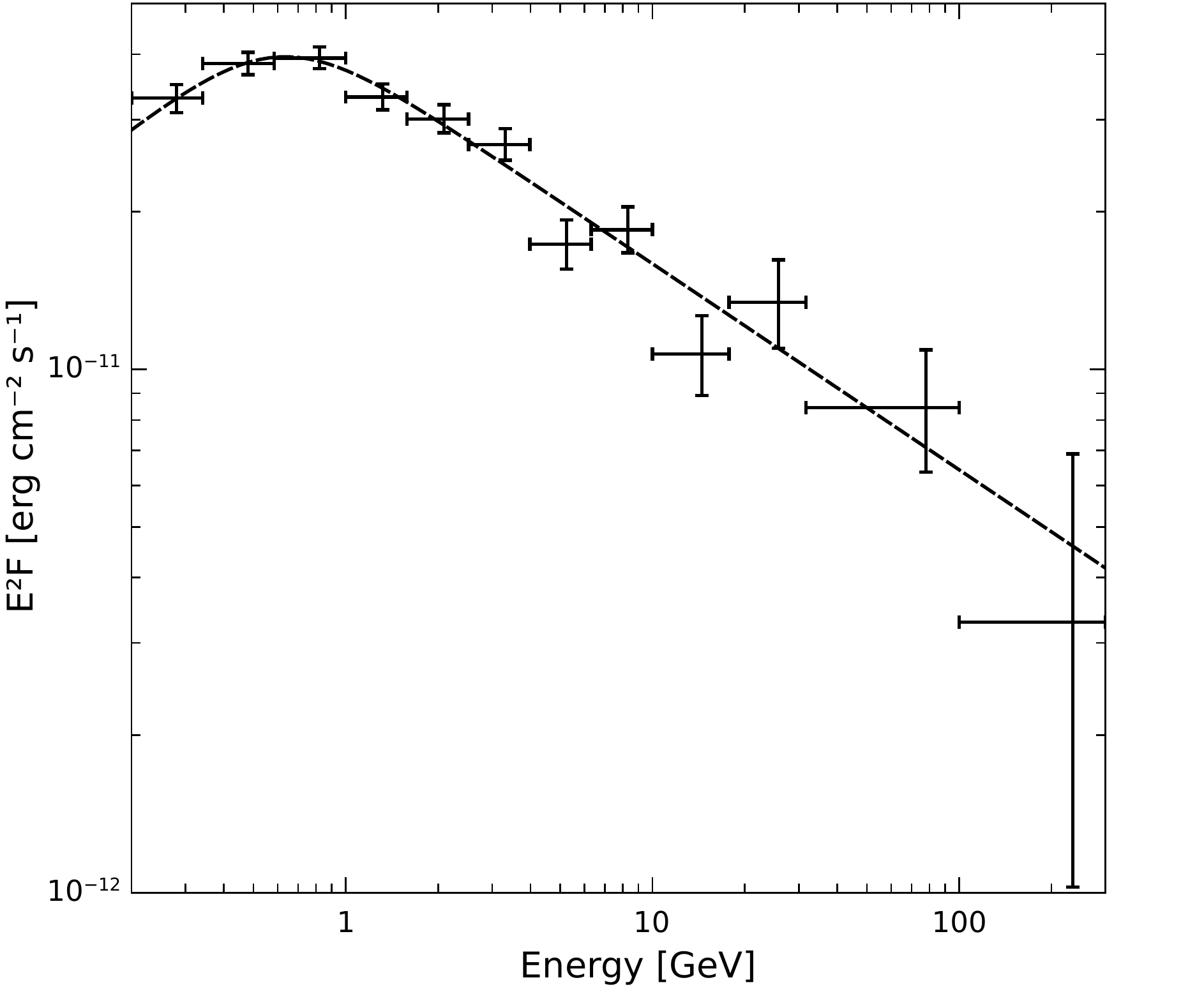}
\end{minipage}
\begin{minipage}{0.49\textwidth}
\includegraphics[width=\textwidth]{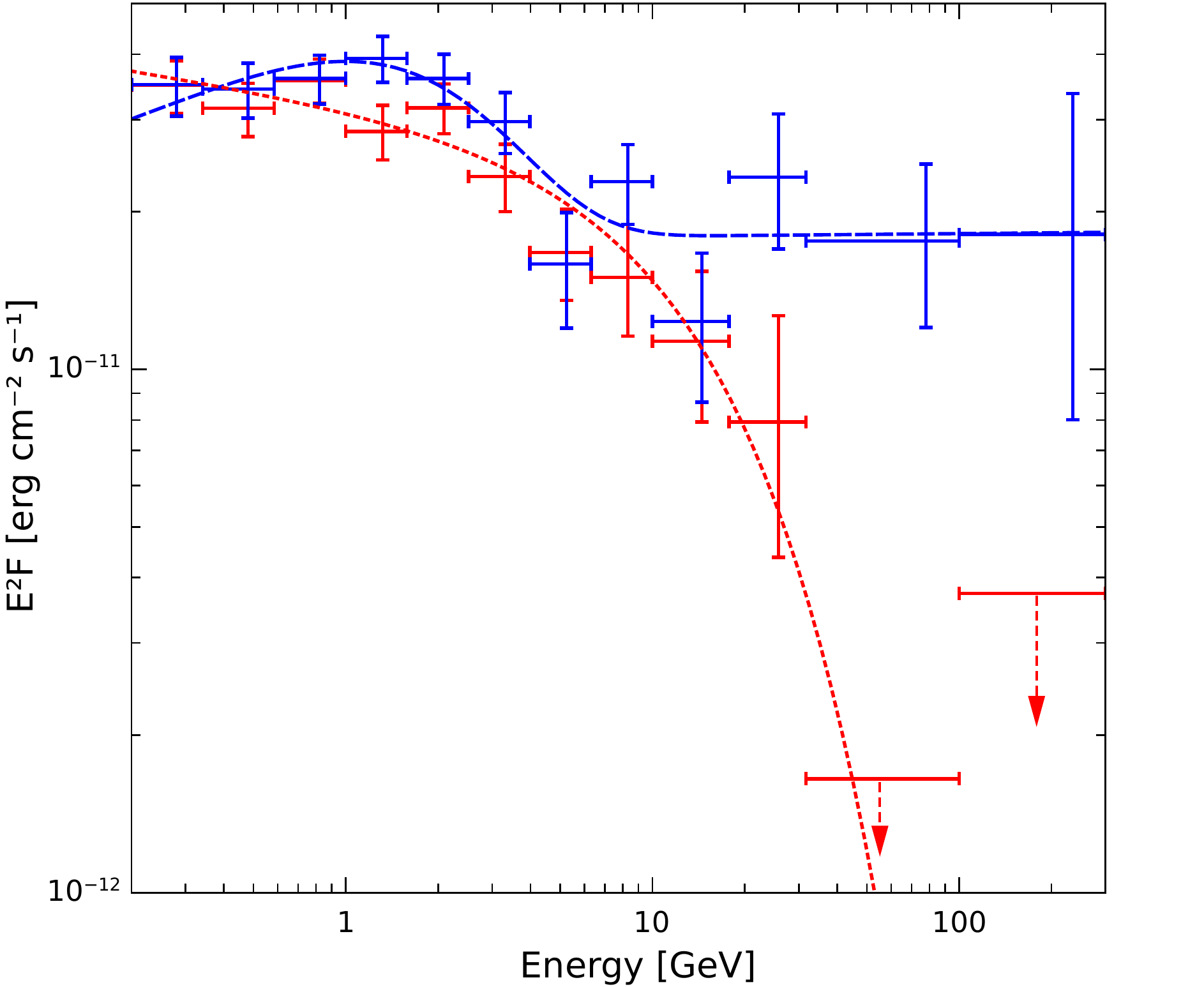}
\end{minipage}
\caption{Spectral energy distribution of $\eta$~Carinae obtained by
likelihood analysis (points). The error bars are of 1$\sigma$ type and only reflect the statistical uncertainties. The upper limits (represented by arrows) have been determined such that the difference of the logarithmic likelihood values (with and without an additional trial point source at the indicated flux value) correspond to 1$\sigma$. The line depicts the best fit of the spectral model that best describing the spectra.
Left: representing the full orbit dataset.
Right: representing 500 days centered on the periastron passage (blue) and the apastron passage (red). \label{spec}}
\end{figure*}
The energy spectrum of $\eta$~Carinae was determined by applying the same maximum-likelihood fitting technique as above to individual energy bins that were chosen to be consistent with earlier studies \citep{Reitberger2012}.

Figure \ref{spec} (left) shows the spectrum for the total dataset of 2024 days and is best described by a SmoothBrokenPowerLaw (SBPL) function of the form
\begin{equation} \frac{dN}{dE}=A\left(\frac{E}{E_0}\right)^\alpha\left(1+\left(\frac{E}{E_b}\right)^\frac{\alpha-\beta}{\epsilon}\right)^{-\epsilon} ,\end{equation}
with the fit parameters as given in Table \ref{fits}.

Quite different spectra are obtained for a period of 500 days centered on the apastron and periastron passages (phase 0.5 and 0 or 1), see Fig. \ref{spec} (right). The choice of these two time intervals is a compromise between the objective to obtain representative spectra for the time of lowest and highest stellar separation and the need for sufficiently high (and roughly equal) statistics to allow for comparison of the two. Owing to the early occurrence of the periastron passage shortly after the beginning of LAT observations in 2008, the respective interval includes data from the beginning of the full-orbit dataset ($\sim$410 days) and the end ($\sim$90 days). The two distinct spectra indicate that the full-orbit spectra cannot be viewed as representative for each phase of the orbit. Instead, it is the average of quite distinct orbital states. This predominantly concerns the high-energy regime above $\sim$20 GeV where the difference between the apastron and periastron spectra is evident. For the energy bin 32 to 100 GeV (second from right) the periastron value of the differential flux is a factor of 10 higher than the upper limit determined at apastron.
The fit to the periastron spectrum is a CPL+PL function (following the approach in \cite{Reitberger2012}), which serves best to describe the two-component spectral shape. An alternative fit to the periastron spectrum by a CPL model yields an unsatisfactory presentation of the data ($\chi^2/dof$ $>$ 2.2) and was rejected against the CPL+PL with more than 2$\sigma$. The F-test between both spectral models gives a value of 3.87, which rejects the null hypothesis that both models fit equally well at 94$\%$ confidence.

For the apastron spectrum the second component at high energies disappears along with minor less significant alterations at the low-energy band. The favored spectral model now is a simple CPL function. The respective parameters are listed in Table \ref{fits}.

\begin{table}
\center
\begin{tabular}{c c c c c}
\hline \hline
 parameter & SBPL & CPL+PL & CPL \\
 & full-orbit & periastron & apastron\\
\hline
$A$ & 38 $\pm$ 50 &  25.2 $\pm$ 9.8 & 20.4 $\pm$ 1.2 & 1$0^{-12}$ MeV$^{-1}$cm$^{-2}$s$^{-1}$ \\
$\alpha$ & 1.54 $\pm$ 0.89 & 2.0 $\pm$ 0.1 & 2.09 $\pm$ 0.06 \\
$B$ & - & 11.1 $\pm$ 4.4 & - & 1$0^{-12}$ MeV$^{-1}$cm$^{-2}$s$^{-1}$ \\
$\beta$ & 2.39 $\pm$ 0.06 & 1.35 $\pm$ 0.57 & - & \\
$E_b$ & 0.60 $\pm$ 0.56 & 1.5 $\pm$ 1.0 & 16.9 $\pm$ 6.1 & GeV \\
$\epsilon$ & 0.28 $\pm$ 0.17 & - & - &  \\
\hline
\end{tabular}
\caption{Best-fit parameters for functions (4), (1+3), and (3) fitted to the full-orbit, periastron and apastron spectra. The scale energy $E_0$ was set to 1 GeV.}
\label{fits}
\end{table}

We studied the SEDs of the nearby pulsar PSR J1048$-$5832 as a cross-check (see comparison in \cite{Reitberger2012} for the same 500-day time intervals)and did not find any significant variations in its spectrum. Measured data points overlap within the 1$\sigma$ error bars.

\subsection{Temporal analysis}
\label{TA}

We now investigate the flux time history of the observed $\gamma$-ray emission. As in the previous section, we analyzed the two adjacent energy intervals 0.2 to 10 GeV and 10 to 300 GeV.

\subsubsection{Flux studies in the low-energy band (0.2 to 10 GeV)}

In Fig. \ref{flux}, left, we show the flux time history of $\eta$~Carinae for the energy band 0.2 to 10 GeV as obtained by likelihood analysis. The dataset was divided into ten consecutive intervals of equal exposure. The durations of each bin in real time are listed in Table \ref{intervals}.

For each time bin we fit a CPL function (all parameters free) to the data. Taking the average flux as determined for the whole dataset (see Sect. \ref{flux}) as the hypothesis for a non-varying source, a reduced $\chi^2$ of 0.78 for 9 degrees of freedom was found. This corresponds to the observation in SEDs in Sect. \ref{SA}, where no significant variability was detected in the low-energy part of the spectrum. Owing to the larger statistics, the relative errors on the flux values in the low-energy band are substantially smaller than those in the high-energy band, which, in principle, would imply a higher receptivity for establishing flux variability than in the high-energy band.

\begin{table}
\center
\begin{tabular}{c c c}
\hline \hline
 bin & $\Delta$t (days) & $\Delta$t (days)\\
 & 0.2 to 10 GeV & 10 to 300 GeV\\
\hline
1 & 209 & 216 \\
2 & 211 & 214\\
3 & 195 & 192 \\
4 & 197 & 197 \\
5 & 197 & 192 \\
6 & 198 & 198  \\
7 & 195 & 194  \\
8 & 203 & 202  \\
9 & 201 & 200 \\
10& 219 & 219  \\
\hline
\end{tabular}
\caption{Time bins of equal exposure as used for the temporal analysis.}
\label{intervals}
\end{table}

\subsubsection{Flux studies in the high-energy band (10 to 300 GeV)}
\label{tempo}

In Fig. \ref{flux}, right, we show the corresponding flux time history for the high-energy band. For each time bin a PL function (all parameters free) was fit to the data. Again, ten time bins of equal exposure (as listed in Table \ref{intervals}) were considered. The variability is far more pronounced than in the low-energy band. The fluxes for the first two bins---roughly centered on the periastron passage---are about a factor of 3.5 higher than the fluxes for the subsequent four bins that cover the orbital phases before the apastron passage. After apastron, the flux shows a roughly gradual increase until it reaches the level of the previous periastron flux shortly before the next periastron passage.
A $\chi^2$-test (reduced $\chi^2$ of 2.00 for nine degrees of freedom) shows that a null-hypothesis of observing a constant average flux (as determined in Sect. \ref{flux}) is disfavored at the level of 2.5$\sigma$.
The X-ray light-curve does not resemble sinusodial modulation, but has a spike plus deep absorption trough around periastron \citep{Mottat09}, therefore we devised a different statistical test to quantify the indicated orbital modulation.
Under the assumption that orbital modulation is marked by highest $\gamma$-ray flux around periastron, as formulated in \cite{Reitberger2012} on the basis of three fifths of the ten light-curve (lc) bins in Fig. \ref{flux}, the complete lc over the full orbital phase can now be investigated.To avoid introducing biases, we did not select lc bins individually that may or may not relate uniquely to a periastron high-flux condition. Instead, we tested a multitude of lc bin combinations and subsequently corrected the obtained result for trials. For this we considered all fractions of the lc comprising two, three, and four consecutive lc bins that include the periastron passage and that are roughly symmetric with respect to phase zero.

The $\gamma$-ray flux observed over the periastron passage differs by 4.5$\sigma$ (post-trial significance) from the rest of the 10 to 300 GeV $\gamma$-ray light curve.

\begin{figure*}

\center
\begin{minipage}{0.49\textwidth}
\centering
\includegraphics[width=\textwidth]{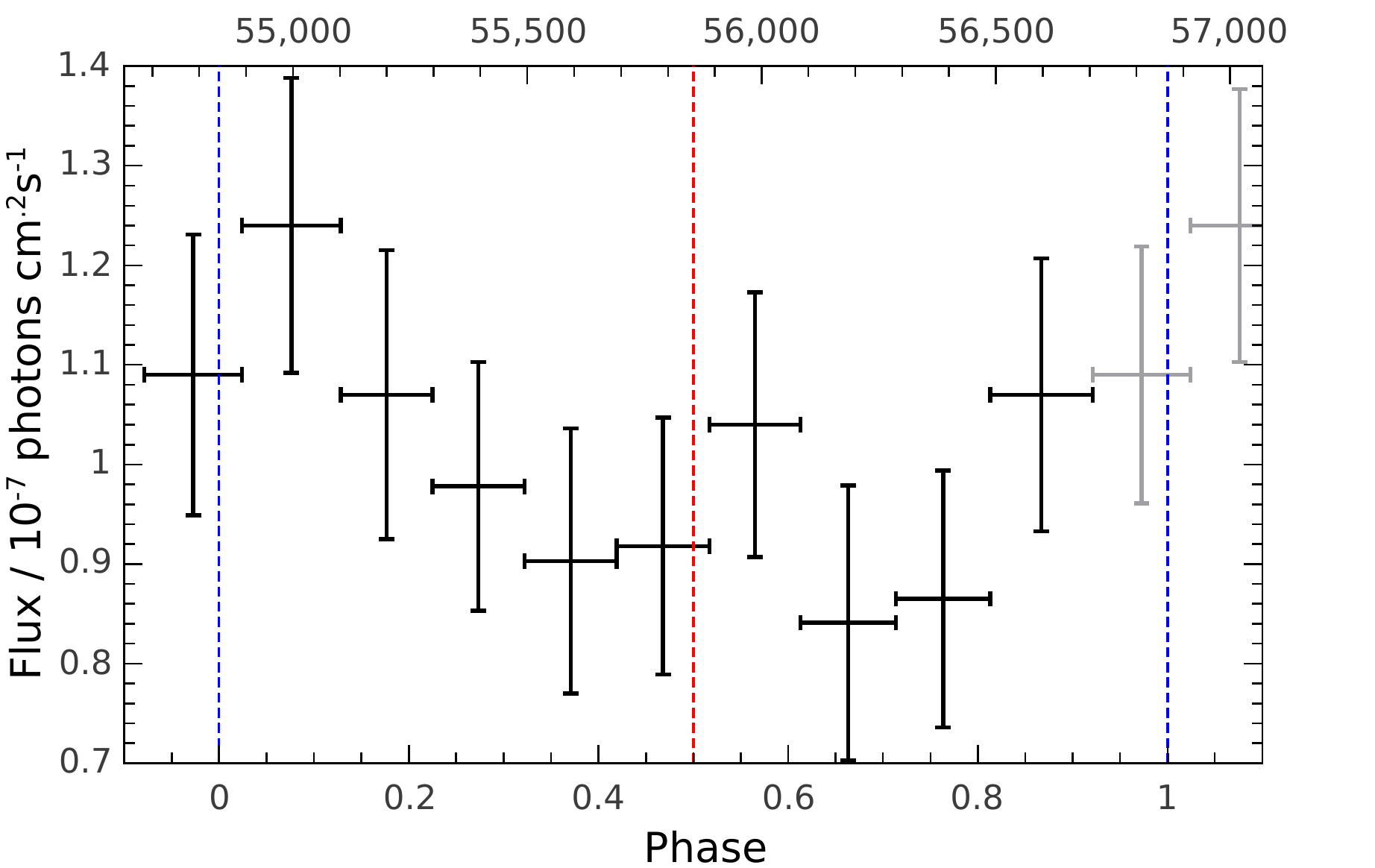}

\end{minipage}
\begin{minipage}{0.49\textwidth}
\centering
\includegraphics[width=\textwidth]{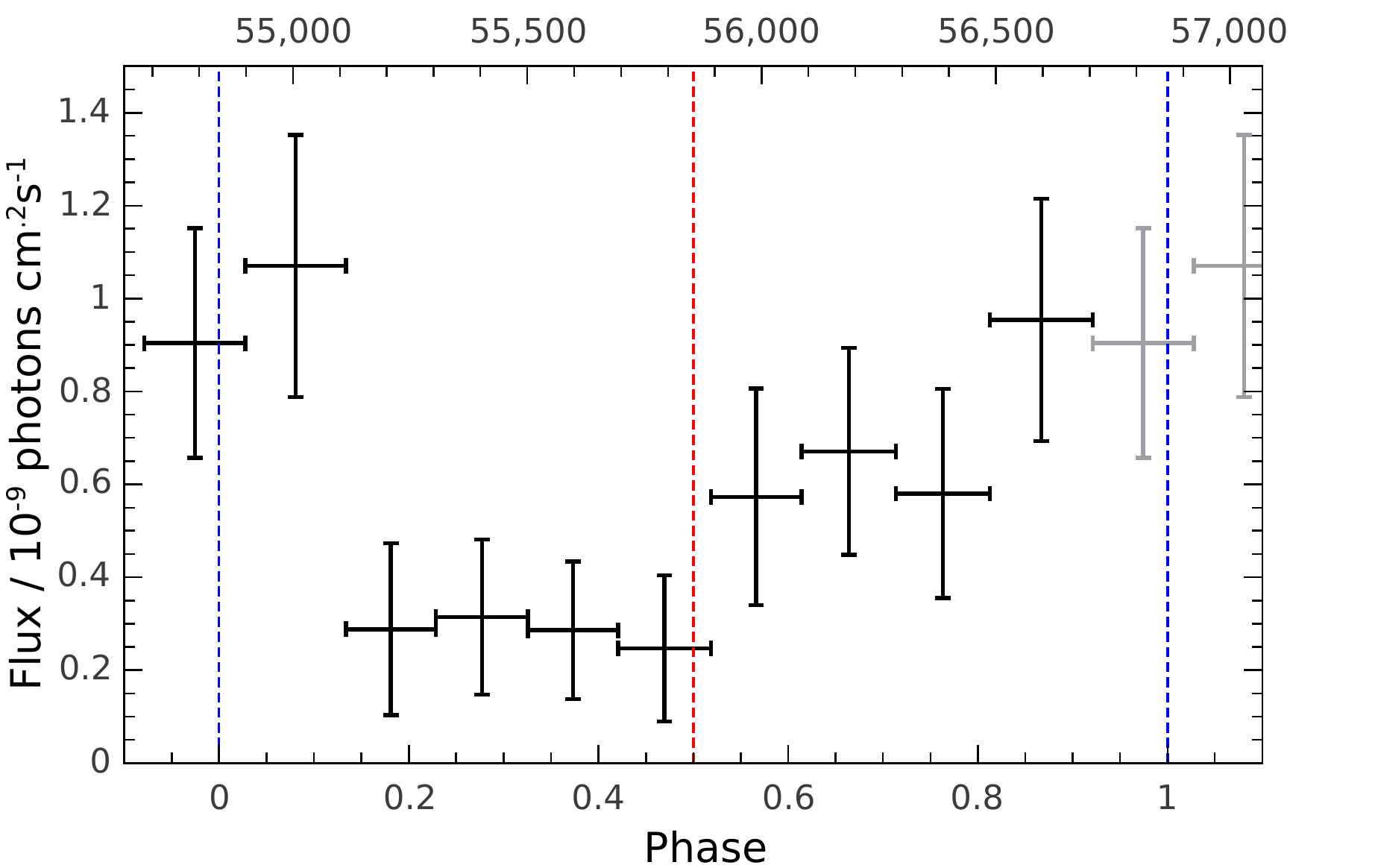}
\end{minipage}

\caption{Left: Flux time history of $\eta$~Carinae in the 0.2 to 10 GeV energy band obtained by likelihood analysis. The lower x-axis is in units of orbital phase of the $\eta$ Carinae binary system, the upper in units of Modified Julian Day (MJD). The flux error bars are of 1$\sigma$ type. The gray data points on the right are the phase-shifted counterparts of the first two data points on the left. The dashed vertical line indicates the time of the periastron passage (blue) and the apastron passage (red) of the $\eta$~Carinae binary system. Right: The same for energies 10 to 300 GeV
.\label{flux} }
\end{figure*}

The significant variability in concordance with the orbital period at energies 10 to 300 GeV agrees with what we reported from the spectra of Sect. \ref{SA} and also with an earlier analysis \citep{Reitberger2012}. There are some differences between the light curves in \cite{Reitberger2012} and those presented here, however. We highlight three aspects before we discuss this: First, this analysis was carried out using newer versions of the LAT datasets, instrument response functions, and background models, which allows for a more accurate analysis now. Second, we changed from lc bins that are equally spaced in time to those based on equal exposure, now aimed for minimizing effects from exposure differences between individual lc bins. Last, even then the differences are at low significance. Comparing the light curves nevertheless, there is a mild indication that including the exact time of the periastron passage tends to produce a mildly lower gamma-ray flux than in near periastron bins that do not. This hint neededs to be studied in more detail on short orbital phase intervals. Because the Fermi data do not yield significant detections on distinctively shorter lc bins, it might be investigated using model predictions.

\section{Discussion and summary}
\label{Disc}
We have shown that flux variability at energies above 10 GeV prevails throughout the complete first full-orbit dataset of a $\gamma$-ray source at the position of $\eta$ Car. Using a new dataset (Pass7 reprocessed), revised and updated IRFs (P7REP$\_$SOURCE$\_$V15 instead of P7SOURCE$\_$V6), a new source model (based on the 3FGL instead of the 2FGL catalog), as well as updated background models, we confirmed the previously observed \citep[see ][]{Reitberger2012} decrease in $\gamma$-ray flux at energies from 10 to 300 GeV. In addition, we tested the previously predicted increase in flux at high energies that had been expected to occur after the last apastron passage. It has been shown that the flux recently attained levels that resemble
those before the last periastron passage in late 2008. Although caused by nonthermal processes instead of thermal ones, the high-energy $\gamma$-flux light curve above 10 GeV shows some general qualitative resemblance with $\eta$ Carinae's thermal X-ray signal as measured by RXTE \citep{Corcoran05}, related to the fact that both depend on conditions in the binary's WCR.

The full-orbit dataset allowed us to compare the spectra centred on the apastron and periastron passages. They clearly deviate in the high-energy part, which can be modeled by an additional hard PL component emerging around the periastron passage. For energies from 32 to 100 GeV, the observed flux at periastron is more than a factor of 10 higher than the upper limit determined at apastron.

The phase-locked flux changes in the high-energy emission component together with improved spatial agreement with the location of $\eta$ Carinae strengthens the case for a firm identification.

We currently have two scenarios that may account for the spectral variation observed in $\eta$ Carinae in the context of a colliding-wind binary interpretation for high-energy \mbox{$\gamma$-ray} emission \citep{Eichler1993, Reimer2006, Dougherty2006}. The spectral extension emerging at periastron could \citep[as proposed by ][]{Walter} be due to an additional (fully or partly hadronic) $\gamma$-ray emission component. This scenario is also plausible in the light of recent 3D hydrodynamical studies that take into account the dynamics of the WCR and simultaneously solve a transport equation for shock-accelerated particles \citep[see ][]{Reitberger2014,Reitberger2014b}. Although not yet applied to the specific parameters of the $\eta$ Carinae binary system, these models show that considerable variations---even a transition from lepton- to hadron-dominated $\gamma$-ray emission---occur with changing stellar separation in a typical colliding-wind binary system. It remains to be seen what the enhanced 3D hydrodynamic models predict for the $\eta$ Carinae binary systems when they
are applied to its specific stellar and orbital parameters. However, a comparison of model predictions and observational $\gamma$-ray data for this peculiar binary system is extremely challenging because of the complexity and turbulence of the wind collision region \citep[see e.g., ][]{Madura2013}.

An alternative to postulating a second emission component emerging around periastron relates to phase-modulated strong photon-photon absorption in the complex radiation fields in the vicinity of $\eta$ Carinae. As discussed in \cite{Reitberger2012}, the observed spectral variation might be explained by the presence of hot shocked gas in the WCR or by spatially extended X-ray emission components surrounding the system and the ensuing photon-photon absorption. Different properties of an absorber at periastron and apastron conditions could still respond to the observed spectra, rendering such a
scenario less likely, although not firmly excluding it.

Ongoing sophisticated modeling efforts may provide more insights into these questions.  The high-energy $\gamma$-ray flux up to 300 GeV around periastron raises expectations that
it might be detected with the IACT instrument \textit{H.E.S.S 2}, at least around the periastron passage.

\section{Acknowledgements}
\begin{acknowledgements}
The \textit{Fermi} LAT Collaboration acknowledges generous ongoing support
from a number of agencies and institutes that have supported both the
development and the operation of the LAT as well as scientific data analysis.
These include the National Aeronautics and Space Administration and the
Department of Energy in the United States, the Commissariat \`a l'Energie Atomique
and the Centre National de la Recherche Scientifique / Institut National de Physique
Nucl\'eaire et de Physique des Particules in France, the Agenzia Spaziale Italiana
and the Istituto Nazionale di Fisica Nucleare in Italy, the Ministry of Education,
Culture, Sports, Science and Technology (MEXT), High Energy Accelerator Research
Organization (KEK) and Japan Aerospace Exploration Agency (JAXA) in Japan, and
the K.~A.~Wallenberg Foundation, the Swedish Research Council and the
Swedish National Space Board in Sweden.\\
Additional support for science analysis during the operations phase is gratefully
acknowledged from the Istituto Nazionale di Astrofisica in Italy and the Centre National d'\'Etudes Spatiales in France.\\
The publication is supported by the Austrian Science Fund (FWF).
\end{acknowledgements}


\bibliographystyle{aa}

\begin{thebibliography}{20}
\expandafter\ifx\csname natexlab\endcsname\relax\def\natexlab#1{#1}\fi

\bibitem[{{Abdo} {et~al.}(2010){Abdo}, {Ackermann}, {Ajello}, {Allafort},  {Baldini}, {Ballet}, {Barbiellini}, {Bastieri}, {Bechtol}, {Bellazzini},
  {Berenji}, {Blandford}, {Bonamente}, {Borgland}, {Bouvier}, {Brandt},
  {Bregeon}, {Brez}, {Brigida}, {Bruel}, {Buehler}, {Burnett}, {Caliandro},
  {Cameron}, {Caraveo}, {Carrigan}, {Casandjian}, {Cecchi}, {{\c C}elik},  {Chaty}, {Chekhtman}, {Cheung}, {Chiang}, {Ciprini}, {Claus}, {Cohen-Tanugi},
  {Cominsky}, {Conrad}, {Dermer}, {de Palma}, {Digel}, {Silva}, {Drell},
  {Dubois}, {Dumora}, {Favuzzi}, {Fegan}, {Ferrara}, {Frailis}, {Fukazawa},
  {Fusco}, {Gargano}, {Gehrels}, {Germani}, {Giglietto}, {Giordano}, {Godfrey},
  {Grenier}, {Grondin}, {Grove}, {Guillemot}, {Guiriec}, {Hadasch}, {Hanabata},
  {Harding}, {Hayashida}, {Hays}, {Hill}, {Horan}, {Hughes}, {Itoh}, {Jackson},
  {J{\'o}hannesson}, {Johnson}, {Johnson}, {Kamae}, {Katagiri}, {Kataoka},
  {Kerr}, {Kn{\"o}dlseder}, {Kuss}, {Lande}, {Latronico}, {Lee},
  {Lemoine-Goumard}, {Livingstone}, {Llena Garde}, {Longo}, {Loparco},
  {Lovellette}, {Lubrano}, {Makeev}, {Mazziotta}, {McEnery}, {Mehault},
  {Michelson}, {Mitthumsiri}, {Mizuno}, {Moiseev}, {Monte}, {Monzani},
  {Morselli}, {Moskalenko}, {Murgia}, {Nakamori}, {Naumann-Godo}, {Nolan},
  {Norris}, {Nuss}, {Ohsugi}, {Okumura}, {Omodei}, {Orlando}, {Ormes}, {Ozaki},
  {Panetta}, {Parent}, {Pelassa}, {Pepe}, {Pesce-Rollins}, {Piron}, {Porter},
  {Rain{\`o}}, {Rando}, {Razzano}, {Reimer}, {Reimer}, {Reposeur}, {Rodriguez},
  {Romani}, {Roth}, {Sadrozinski}, {Sander}, {Saz Parkinson}, {Scargle},
  {Sgr{\`o}}, {Siskind}, {Smith}, {Smith}, {Spandre}, {Spinelli}, {Strickman},
  {Suson}, {Takahashi}, {Takahashi}, {Tanaka}, {Thayer}, {Thayer}, {Thompson},
  {Tibaldo}, {Tibolla}, {Torres}, {Tosti}, {Tramacere}, {Uchiyama}, {Usher},
  {Vandenbroucke}, {Vasileiou}, {Vilchez}, {Vitale}, {Waite}, {Wallace},
  {Wang}, {Winer}, {Wood}, {Yang}, {Ylinen}, \& {Ziegler}}]{Takapaper}
{Abdo}, A.~A., {Ackermann}, M., {Ajello}, M., {et~al.} 2010, ApJ, 723, 649

\bibitem[{{Acero} {et~al.}(2015) {The Fermi-LAT Collaboration}}] {3FGL}
{Acero}, F., {Ackermann}, M., {Ajello}, M., {et~al.} 2015, arXiv:1501.02003


\bibitem[{{Ackermann} {et~al.}(2012){Ackermann}, {Ajello}, {Albert},
  {Allafort}, {Atwood}, {Axelsson}, {Baldini}, {Ballet}, {Barbiellini},
  {Bastieri}, {Bechtol}, {Bellazzini}, {Bissaldi}, {Blandford}, {Bloom},
  {Bogart}, {Bonamente}, {Borgland}, {Bottacini}, {Bouvier}, {Brandt},
  {Bregeon}, {Brigida}, {Bruel}, {Buehler}, {Burnett}, {Buson}, {Caliandro},
  {Cameron}, {Caraveo}, {Casandjian}, {Cavazzuti}, {Cecchi}, {{\c C}elik},
  {Charles}, {Chaves}, {Chekhtman}, {Cheung}, {Chiang}, {Ciprini}, {Claus},
  {Cohen-Tanugi}, {Conrad}, {Corbet}, {Cutini}, {D'Ammando}, {Davis}, {de  Angelis}, {DeKlotz}, {de Palma}, {Dermer}, {Digel}, {Silva}, {Drell},
  {Drlica-Wagner}, {Dubois}, {Favuzzi}, {Fegan}, {Ferrara}, {Focke}, {Fortin},
  {Fukazawa}, {Funk}, {Fusco}, {Gargano}, {Gasparrini}, {Gehrels}, {Giebels},
  {Giglietto}, {Giordano}, {Giroletti}, {Glanzman}, {Godfrey}, {Grenier},  {Grove}, {Guiriec}, {Hadasch}, {Hayashida}, {Hays}, {Horan}, {Hou}, {Hughes},
  {Jackson}, {Jogler}, {J{\'o}hannesson}, {Johnson}, {Johnson}, {Johnson},
  {Kamae}, {Katagiri}, {Kataoka}, {Kerr}, {Kn{\"o}dlseder}, {Kuss}, {Lande},
  {Larsson}, {Latronico}, {Lavalley}, {Lemoine-Goumard}, {Longo}, {Loparco},
  {Lott}, {Lovellette}, {Lubrano}, {Mazziotta}, {McConville}, {McEnery},
  {Mehault}, {Michelson}, {Mitthumsiri}, {Mizuno}, {Moiseev}, {Monte},
  {Monzani}, {Morselli}, {Moskalenko}, {Murgia}, {Naumann-Godo}, {Nemmen},
  {Nishino}, {Norris}, {Nuss}, {Ohno}, {Ohsugi}, {Okumura}, {Omodei},
  {Orienti}, {Orlando}, {Ormes}, {Paneque}, {Panetta}, {Perkins},
  {Pesce-Rollins}, {Pierbattista}, {Piron}, {Pivato}, {Porter}, {Racusin},
  {Rain{\`o}}, {Rando}, {Razzano}, {Razzaque}, {Reimer}, {Reimer}, {Reposeur},
  {Reyes}, {Ritz}, {Rochester}, {Romoli}, {Roth}, {Sadrozinski}, {Sanchez},
  {Saz Parkinson}, {Sbarra}, {Scargle}, {Sgr{\`o}}, {Siegal-Gaskins},
  {Siskind}, {Spandre}, {Spinelli}, {Stephens}, {Suson}, {Tajima}, {Takahashi},
  {Tanaka}, {Thayer}, {Thayer}, {Thompson}, {Tibaldo}, {Tinivella}, {Tosti},
  {Troja}, {Usher}, {Vandenbroucke}, {Van Klaveren}, {Vasileiou}, {Vianello},
  {Vitale}, {Waite}, {Wallace}, {Winer}, {Wood}, {Wood}, {Wood}, {Yang}, \&
  {Zimmer}}]{IRFpaper}
{Ackermann}, M., {Ajello}, M., {Albert}, A., {et~al.} 2012, \apjs, 203, 4
\bibitem[{{Bednarek} \& {Pabich}(2011)}]{Bednarek2011}
{Bednarek}, W. \& {Pabich}, J. 2011, \aap, 530, A49

\bibitem[{{Corcoran}(2005)}]{Corcoran05}
{Corcoran}, M.~F. 2005, \aj, 129, 2018

\bibitem[{{Moffat} \& {Corcoran}(2009)}]{Mottat09}
{Moffat}, A.~F.~J. \& {Corcoran}, M.~F. 2009, \apj, 707, 693

\bibitem[{{Eichler} \& {Usov}(1993)}]{Eichler1993}
{Eichler}, D. \& {Usov}, V. 1993, \apj, 402, 271

\bibitem[{{Farnier} {et~al.}(2011){Farnier}, {Walter}, \& {Leyder}}]{Walter}
{Farnier}, C., {Walter}, R., \& {Leyder}, J.-C. 2011, \aap, 526, A57

\bibitem[{{Abramowski} {et~al.}(2012){HESS Collaboration},
  {Abramowski}, {et~al.}}]{Hess}
{Abramowski, A., Acero, F., Aharonian, F., et al. (H.E.S.S. Collaboration)} 2012, MNRAS, 424, 128

\bibitem[{{Leyder} {et~al.}(2008){Leyder}, {Walter}, \& {Rauw}}]{Leyder}
{Leyder}, J.-C., {Walter}, R., \& {Rauw}, G. 2008, \aap, 477, L29

\bibitem[{{Madura} {et~al.}(2013){Madura}, {Gull}, {Okazaki}, {Russell},
  {Owocki}, {Groh}, {Corcoran}, {Hamaguchi}, \& {Teodoro}}]{Madura2013}
{Madura}, T.~I., {Gull}, T.~R., {Okazaki}, A.~T., {et~al.} 2013, \mnras, 436,
  3820

\bibitem[{Mattox {et~al.}(1996)}]{Mattox}
Mattox, J. {et~al.} 1996, ApJ, 461, 396

\bibitem[{{Nolan} {et~al.}(2012){Nolan}, {Abdo}, {Ackermann}, {Ajello},
  {Allafort}, {Antolini}, {Atwood}, {Axelsson}, {Baldini}, {Ballet}, \&
  et~al.}]{2FGL}
{Nolan}, P.~L., {Abdo}, A.~A., {Ackermann}, M., {et~al.} 2012, ApJS, 199, 31

\bibitem[{Ohm {et~al.}(2010)}]{Ohm}
Ohm, S. {et~al.} 2010, ApJ Letters, 718, L161

\bibitem[{{Pittard} \& {Dougherty}(2006)}]{Dougherty2006}
{Pittard}, J.~M. \& {Dougherty}, S.~M. 2006, MNRAS, 372, 801

\bibitem[{Reimer {et~al.}(2006)}]{Reimer2006}
Reimer, A. {et~al.} 2006, MNRAS, 644, 1118

\bibitem[{{Reitberger} {et~al.}(2014{\natexlab{a}}){Reitberger}, {Kissmann},
  {Reimer}, \& {Reimer}}]{Reitberger2014b}
{Reitberger}, K., {Kissmann}, R., {Reimer}, A., \& {Reimer}, O.
  2014{\natexlab{a}}, \apj, 789, 87

\bibitem[{{Reitberger} {et~al.}(2014{\natexlab{b}}){Reitberger}, {Kissmann},
  {Reimer}, {Reimer}, \& {Dubus}}]{Reitberger2014}
{Reitberger}, K., {Kissmann}, R., {Reimer}, A., {Reimer}, O., \& {Dubus}, G.
  2014{\natexlab{b}}, \apj, 782, 96

\bibitem[{{Reitberger} {et~al.}(2012){Reitberger}, {Reimer}, {Reimer},
  {Werner}, {Egberts}, \& {Takahashi}}]{Reitberger2012}
{Reitberger}, K., {Reimer}, O., {Reimer}, A., {et~al.} 2012, \aap, 544, A98

\bibitem[{{Sekiguchi} {et~al.}(2009){Sekiguchi}, {Tsujimoto}, {Kitamoto},  {Ishida}, {Hamaguchi}, {Mori}, \& {Tsuboi}}]{Sekiguchi}
{Sekiguchi}, A., {Tsujimoto}, M., {Kitamoto}, S., {et~al.} 2009, \pasj, 61, 629

\bibitem[{Tavani {et~al.}(2009)}]{Agile}
Tavani, M. {et~al.} 2009, ApJ Letters, 698, L142

\end{thebibliography}

\clearpage
\onecolumn

\end{document}